\documentclass[twocolumn,prl]{revtex4}
\usepackage{graphics,graphicx}

\begin{document}
\title{NiSi: New venue for antiferromagnetic spintronics}
\author{P. Ghosh$^{1}$}
\author{J. Guo$^{1,\dag}$}
\author{F. Ye$^2$}
\author{T. Heitmann$^{3}$}
\author{S. Kelley$^{4}$}
\author{A. Ernst$^{5}$}
\author{V. Dugaev$^{6}$}
\author{D. K. Singh$^{1,7,*}$}
\affiliation{$^{1}$Department of Physics and Astronomy, University of Missouri, Columbia, MO, USA}
\affiliation{$^{2}$Oak Ridge National Laboratory, Oak Ridge, TN, USA}
\affiliation{$^{3}$University of Missouri Research Reactor, Columbia, MO, USA}
\affiliation{$^{4}$Department of Chemistry, University of Missouri; Columbia, MO, USA}
\affiliation{$^{5}$Institut for Theoretical Physics, Johannes Kepler University, Linz, Austria}
\affiliation{$^{6}$Department of Physics and Medical Engineering, Rzeszów University of Technology, Rzeszów, Poland}
\affiliation{$^{7}$MU Materials Science and Engineering Institute, Columbia, MO, USA}
\affiliation{$^{*}$email: singhdk@missouri.edu}

\begin{abstract}

\textbf{Envisaging antiferromagnetic spintronics pivots on two key criteria of high transition temperature and tuning of underlying magnetic order using straightforward application of magnetic field or electric current. Here, we show that NiSi metal can provide suitable new platform in this quest. First, our study unveils high temperature antiferromagnetism in single crystal NiSi with $T_{N} \geq 700$ K. Antiferromagnetic order in NiSi is accompanied by the non-centrosymmetric magnetic character with small ferromagnetic component in a-c plane. Second, we find that NiSi manifests distinct magnetic and electronic hysteresis responses to field applications due to the disparity in two moment directions. While magnetic hysteresis is characterized by one-step switching between ferromagnetic states of uncompensated moment, electronic behavior is ascribed to metamagnetic switching phenomena between non-collinear spin configurations. Importantly, the switching behaviors persist to high temperature. The properties underscore the importance of NiSi in the pursuit of antiferromagnetic spintronics.}
 
\end{abstract}

 \maketitle

Antiferromagnetism, originally discovered by Louis Neel more than five decades ago, was explained on the basis of quantum mechanical exchange coupling between ions with opposite spin polarity.\cite{Neel,Nagamiya}  The unconventional mechanism, at that time, elucidated the origin behind the absence of net magnetization in antiferromagnetic materials and introduced the concept of sub-lattice magnetizations. While the zero magnetization in an antiferromagnet minimizes the stray field between magnetic sites, thus maximizing the packing density for magnetic random access memory application, strong exchange coupling (compared to ferromagnetic system) enhances the characteristic frequency of spin dynamics that is faster and stable against the external perturbation.\cite{Zelezny}  These properties are at the core of new push behind spintronics research.\cite{Siddiqui,Baltz,Jungwirth,Yan} Antiferromagnetic (AFM) metal provides facile platform in this endeavor.\cite{Pickett, Wadley, Fukami, Wu, Gomonav} An AFM metal tends to exhibit high electrical and thermal conductivities.\cite{Siddiqui} However, unlike ferromagnetic compounds where the anisotropic magnetoresistance (AMR) is commonly used for reading and writing information in the form of two magnetic polarities, separated by an energy barrier, zero net magnetization and stronger spin dynamics in antiferromagnetic compounds spur weak magnetic field dependence. Thus, the reading of information using AMR is difficult in an antiferromagnetic metal, even though the AMR of relativistic origin has been demonstrated in some cases.\cite{Shick}

More recently, novel approaches of current-induced staggered relativistic field and the spin-orbit torque (among others) have provided new mechanisms to alter the magnetic order parameter in antiferromagnetic metals, such as CuMnAs.\cite{Zelezny,Wadley,Siddiqui} The uncoupled applications of magnetic field at high temperature followed by electric current along certain crystallographic directions at relatively lower temperature was demonstrated to manipulate the magnetic state in AFM metal FeRh.\cite{Marti} Similarly, the concept of exchange coupling as a field-tunable parameter in the artificially engineered composite system of thin ferromagnetic and antiferromagnetic layers is actively pursued to control the magnetization direction in antiferromagnetic component.\cite{Wang,Tang,Duine}

\begin{figure*}
\centering
\includegraphics[width=18. cm]{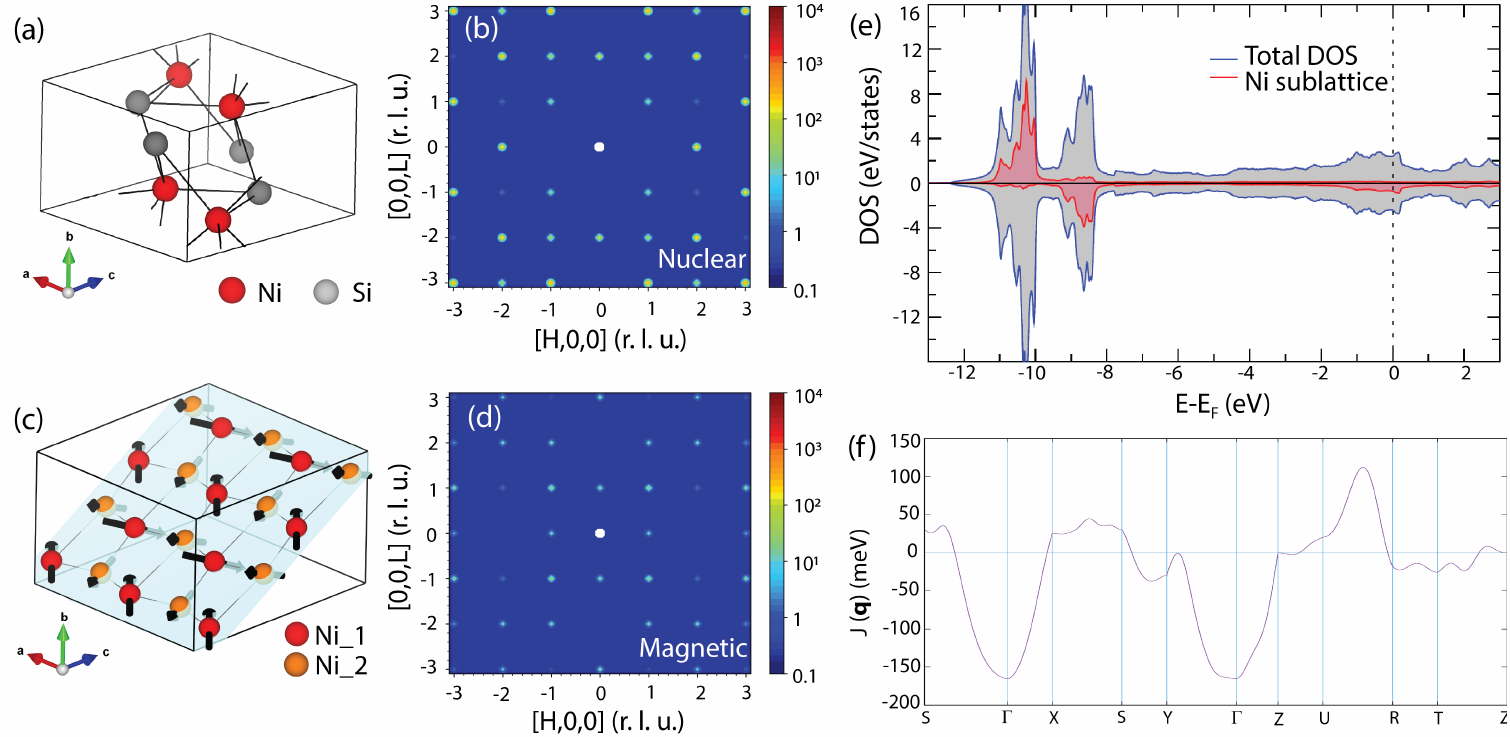} \vspace{-2mm}
\caption{\textbf{Lattice structure of NiSi and spin correlation between Ni atoms.} (a-b) Orthorhombic crystal structure of NiSi. Numerically simulated diffraction profile for the NiSi lattice. (c-d) AFM structure of Ni-ions and the simulated scattering profile, exhibiting many distinct peaks arising solely due to spin correlation between Ni ions, such as (001) and (110). (e-f) Theoretical calculations within the self-interaction correction model exhibit large Ni density of states near the Fermi surface (Fig. e) and maxima in total magnetic energy \textit{J}(\textbf{\textit{q}}) between high symmetry points of \textbf{\textit{R}} and \textbf{\textit{U}} (Fig. f), suggesting AFM propagation vector.} \vspace{-4mm}
\end{figure*}

Unlike the fully compensated AFM metal where the manipulation of magnetic state requires the application of a complex process involving electric current and magnetic field or multiple current sources, a non-compensated AFM metal with small ferromagnetic component provides a rather accessible platform for the reading of information using only magnetic field application. As shown in the case of Mn$_{3}$Si, the anomalous Hall conductivity and magnetoresistance were used to deduce information about the field-induced change in magnetic configuration.\cite{Nakatsuji} Metallic AFM compounds, arguably, exhibit large anomalous Hall effect due to the conjugated operations of time-reversal and space inversion symmetries.\cite{Vries,Zhang,Chen,Nakatsuji} However, there are two benchmark criteria that an AFM metal must satisfy for spintronic applications: it must have high Neel temperature, $T_N$ $>$ 500 K, for the resulting device to be practically usable, and also manifest easy assimilation with the complex integrated circuit.\cite{Samarth} 

Nickel monosilicide (NiSi) fits both criteria. NiSi is a technologically important electronic material. It is used for making electrical contacts in field-effect transistors and nanoelectronic devices.\cite{Robinson,Lavoe,Weber} In contrast to other bi-elemental transition metal systems that are known to manifest rich magnetic and topological phases,\cite{Chubukov, Nakajima, Huber} magnetic order was never reported in NiSi or any Ni-Si compound. Straightforward density functional calculation under the local density approximation (LDA) is also one of the reasons behind the lack of magnetic information in this compound, as it does not yield a magnetic ground state in NiSi.\cite{Dahal} Additionally, previous experimental investigations were focused on powder samples that often averages out finer details. Synthesis of single crystal specimen of NiSi of high chemical purity solves this conundrum. For the first time, our research shows that nickel monosilicide (NiSi) not only possesses uncompensated AFM ground state with a high Neel temperature of $T_N \geq 700$ K but also exhibits remarkable magnetic switching and electronic properties that can be utilized for the spintronics device designs.

\begin{figure*}
\centering
\includegraphics[width=16. cm]{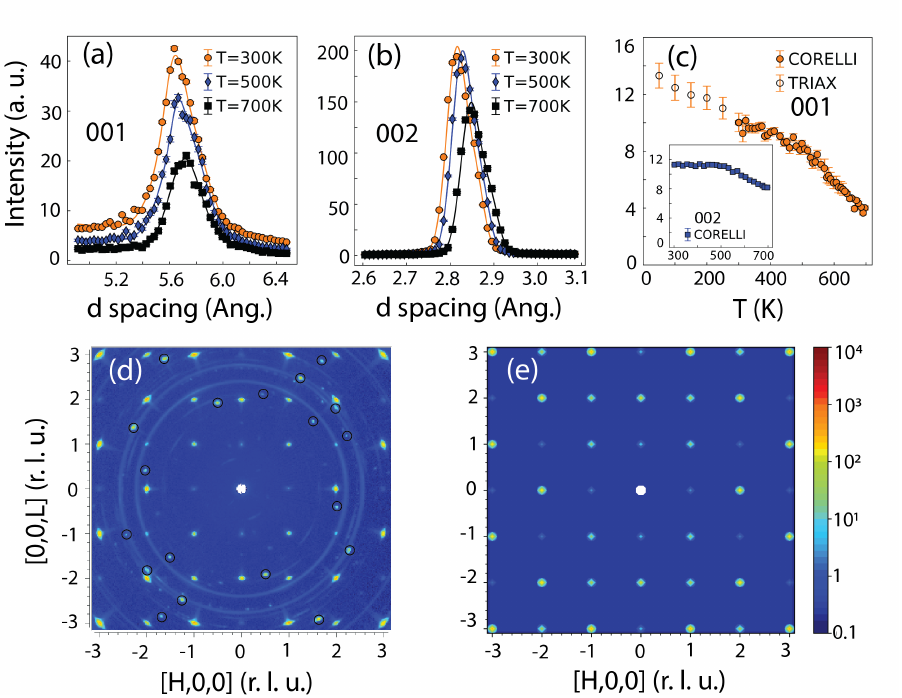} \vspace{-2mm}
\caption{\textbf{Elastic neutron measurements revealing magnetism in NiSi.} (a-b) Representative scans, exhibiting temperature dependence, across (001) and (002) peaks. Experimental data are well-described by resolution limited Gaussian line-shape, suggesting long range magnetic order in NiSi. (c) Plot of integrated intensity as a function of temperature, showing the onset of magnetism at $T \geq 700$ K. Inset shows that the integrated intensity also increases, albeit modestly, at nuclear peak. (d) Detailed mesh scan in [h0l] crystallographic plane at $T = 300$ K. Peaks surrounded by black circles arise due to a small crystallite in the sample, which is not aligned with the main crystal specimen; thus not affecting experimental results. Powder rings are due to aluminum sample can. (e) Numerically simulated scattering profile showing both lattice and magnetic contributions. Error bar represents one standard deviation in all data.} \vspace{-4mm}
\end{figure*}

NiSi crystallizes in the zinc blend-type orthorhombic structure with lattice parameters of $a = 5.178$ \AA, $b = 3.331$ \AA, $c = 5.1624$ \AA. Ni ions occupy the centrosymmetric positions,\cite{Connetable} see Fig. 1a. To understand the origin and nature of magnetism in NiSi, we have performed synergistic theoretical and experimental researches. Unlike the LDA approximation mechanism, theoretical calculations using the self-interaction (SI) correction method reveal an AFM ground state in NiSi.\cite{Luders, Liechtenstein, Dane} We have performed first principle study using a self-consistent full-potential Green’s function method, implemented within the multiple scattering theory. Ni 3d orbitals are almost fully occupied and not spin-polarized. In our theoretical approach, Ni 3d atoms were self-interaction corrected by varying all possible configurations and minimizing the total energy as suggested for monoxides. We find that the Ni 3d ions are occupied in accordance with the Hund’s rule, similar to NiO. Self-interaction corrected Ni 3d states are located mainly at 8-10 eV below the Fermi level. However, a substantial part of the density of states is situated at the Fermi level due to the hybridization with the Si sp-states, see Fig. 1e (also see Fig. S1 in Supporting Information). Nevertheless, a large exchange splitting and the strong localization of the Ni 3d states favor an antiferromagnetic order of Ni magnetic moments. This is confirmed by our total energy calculations. The total energy of the antiferromagnetic stacking per Ni moment is about 40 meV lower than the ferromagnetic configuration.

To theoretically refine magnetic structure of NiSi, we calculate the exchange coupling constants using the magnetic force theorem within the ambit of multiple scattering theorem. Fig. 1f shows the Fourier transformed exchange constants $J(q)$ as a function of the wave vector q. The maximum $J(q)$ corresponds to the minimum in total energy and the corresponding wave vector, associated to magnetic ordering, suggests an AFM order in NiSi. According to our calculations, maximum $J(q)$ is found to be located between the high symmetry points of $\textbf{\textit{U}} = (1/2,0,1/2)$ and $\textbf{\textit{R}} = (1/2,1/2,1/2)$ in the Brillouin zone. It suggests antiferromagnetic order of Ni spins with moment size of 1.12 $\mu_B$, canted in the direction of q. Detailed elastic neutron scattering measurements on single crystal NiSi indeed reveal AFM correlation between Ni-ions with moments symmetrically shifted across the (111) diagonal plane, see Fig. 1c, as discussed in the following paragraphs.

Study of single crystal sample provides crystallographic dependence of the underlying magnetism in a material. We synthesized single crystal specimens of NiSi using the self-flux method via slow cooling. Samples were characterized using powder and Laue X-ray diffraction methods (see Experimental Section). To gain insight into possible magnetism in NiSi, we have performed detailed elastic neutron scattering measurements on a 0.14 g single crystal sample on CORELLI and TRIAX spectrometers at SNS-ORNL and MURR, respectively (see Experimental Section for detail about neutron scattering measurements). Experimental data reveal temperature dependent diffraction peaks as the sample is cooled from $T \sim$ 700 K. We show the representative scans at two q-vectors, \boldmath{$q_1$} = (001) rlu and \boldmath{$q_2$} = (002) rlu, in Fig. 2a-b. \boldmath{$q_1$} and \boldmath{$q_2$} wave-vectors represent distinct magnetic and lattice positions, as shown in Fig. 1b and 1d. The integrated intensities as a function of temperature are plotted in Fig. 2c. While the intensity increases significantly as temperature decreases at \boldmath{$q_1$} vector, indicating the onset of magnetism at \textit{T} $\geq$ 700 K, a modest increment is also detected in peak intensity at \boldmath{$q_2$} wave vector. The transition temperature can be possibly higher as our measurement is limited to \unboldmath{$T = 700$} K.

To deduce the microscopic nature of Ni spin correlation, a detailed mesh scan, obtained at \textit{T} = 300 K, is shown in Fig. 2d (see Fig. S2 for mesh scans in (0kl) and (hk0) scattering planes). Magnetic peak intensity does not change beyond 2$\sigma$ below \textit{T} = 300 K in Fig. 2c. So, we chose this temperature for spin structure refinement. In addition to distinct magnetic and lattice peaks, some peaks (such as (h0h) wave vectors) have both lattice and magnetic contributions. However, magnetic contribution outweighs the lattice at smaller q value. The spin structure is determined by performing magnetic symmetry analysis using FullProf Suite.\cite{Rodriguez} The best fit to experimental data is obtained for the magnetic moment arrangement shown in fig. 1c where oppositely polarized Ni spins, forming AFM configuration, cant out of (111) diagonal plane (see the section on 'Magnetic Structure Refinement' in Supporting Information for more detail). NiSi is not known to exhibit strong uniaxial anisotropy. Therefore, such a non-collinear spin configuration is not surprising. Also, the strength of Ni moment varies on two sites: $m_2$ = 1.58(05) $\mu_{B}$ and $m_2$ = 1.19(04) $\mu_B$. It suggests the non-centrosymmetric character of magnetic unit cell. In addition to the disparity in ordered moments with non-conforming canting on two Ni sites, analysis of change in nuclear peak intensity at \textit{T} = 300 K (Fig. 2c) reveals a non-compensated moment (ferromagnetic (FM) component) per unit cell of magnitude $\sim 0.08(02) \mu_B$ in a-c plane (nominally), given by the polar angles of $\theta = 38.4^{\circ}$ and $\phi = 15.4^{\circ}$. The uncompensated FM moment accompanies the AFM correlation of Ni ions. The numerically simulated pattern of combined magnetic and lattice structure factors, shown in Fig. 2e, depicts excellent agreement with experimental data.

The magnetic correlation of Ni spins can be useful for spintronic applications, as inferred previously in AFM metals.\cite{Siddiqui} The disparity in antiferromagnetic and the uncompensated ferromagnetic moment directions can exhibit different responses to magnetic field applications. To explore the practical implication of this important property of NiSi, we have performed magnetic and Hall probe measurements for different magnetic field application directions. For the tuning of FM component, nominally in the a-c plane, magnetic field is applied along the b-axis. In Fig. 3a, we show the q-independent magnetic measurements at two characteristic temperatures of \textit{T} = 350 K and \textit{T} = 20 K. Magnetic field application perpendicular to the uncompensated moment in a-c plane will induce maximum torque (\text{\boldmath{$\tau$}} = \textbf{\textit{M}} \text{\boldmath{$\times$}}  \textbf{\textit{H}}), thus can force it to align to field direction as described schematically in Fig. 3b. This is precisely what we see in magnetic measurements where highly symmetric hysteresis loops, accompanied by the one-step switching process, is detected. It is characterized by the sharp transition between two-magnetic states at a switching field of $\mu_{0}|$\textbf{\textit{H}}$| \sim$ 0.08 T. Measurement at low temperature does not affect the switching field value. The absence of any field-induced enhancement in magnetization at higher field hints of a magnetization switching process between the two locally minimum energy FM configurations. AFM material with high Neel temperature tends to exhibit paramagnetic behavior at higher field,\cite{Baranov} which is not the case here. Rather, the bulk magnetization saturates to a modest value of 0.67 $\times 10^{-7}$ Am$^{2}$ at \textit{T} = 350 K, which is comparable to the uncompensated FM moment for the measured sample. The one-step switching transition at moderate field sets the foundation for the development of magnetic random access memory spintronic devices. Perhaps, the switching field can be further reduced in the thin film specimen of NiSi where the magnitude of overall magnetization due to the uncompensated FM moment is expected to be even smaller.

\begin{figure}
\centering
\includegraphics[width=9. cm]{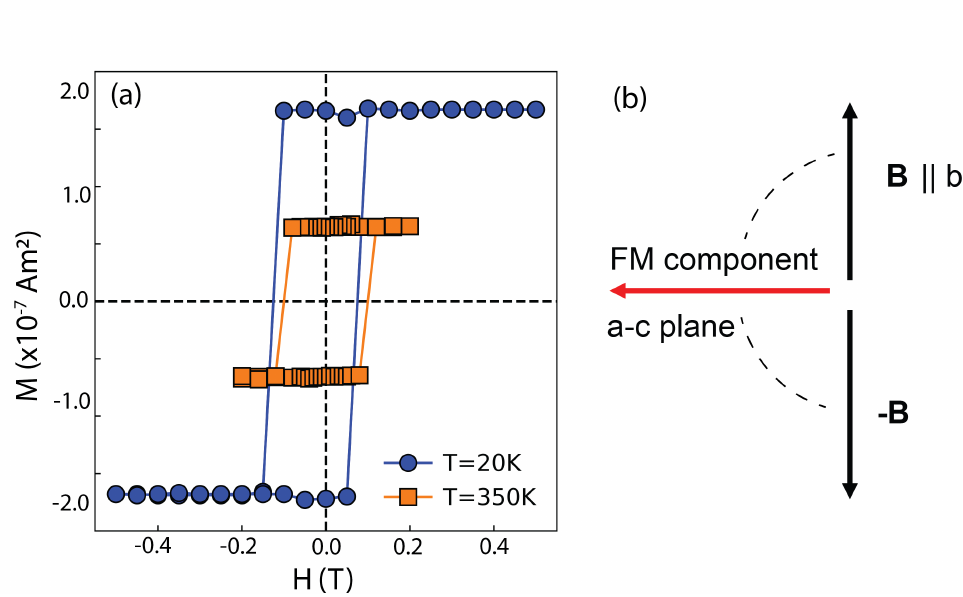} \vspace{-2mm}
\caption{\textbf{Magnetic switching in NiSi.} (a) Bulk magnetometry study reveals a sharp transition between two magnetic states at $H \sim 0.08$ T. The switching occurs due to the field alignment of uncompensated (FM) moment to field direction, depicted schematically in (b).} \vspace{-4mm}
\end{figure}

\begin{figure}
\centering
\includegraphics[width=9. cm]{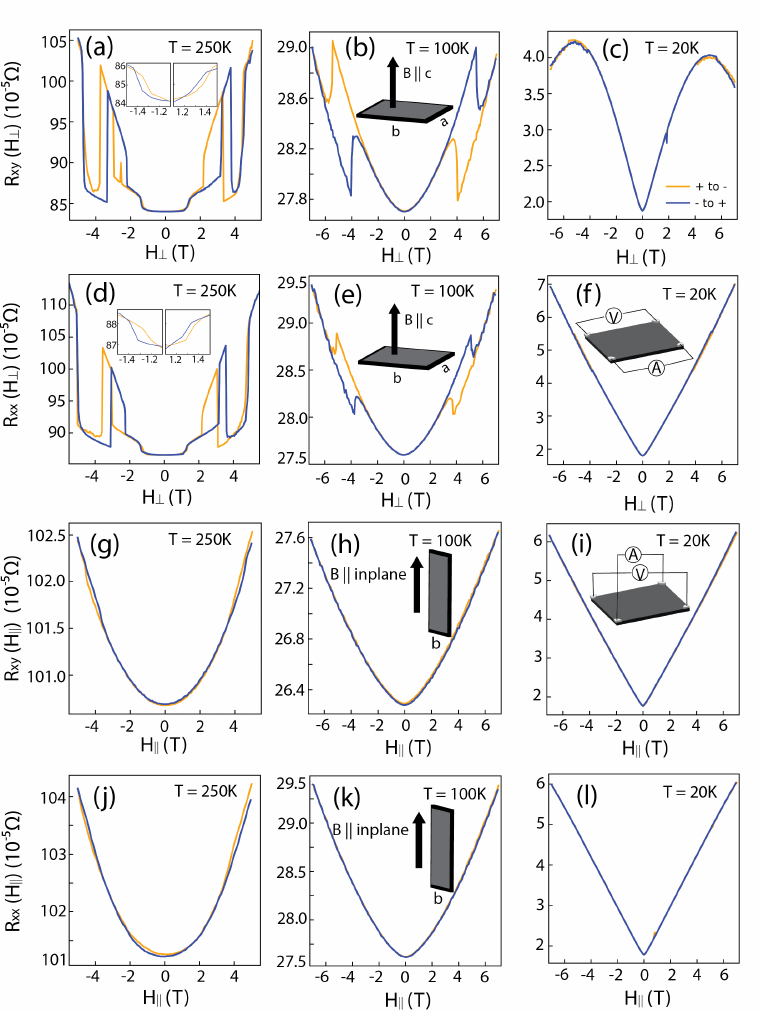} \vspace{-2mm}
\caption{\textbf{Magneto-electronic hysteresis revealed in Hall probe measurements on NiSi.} (a-c) Hall data for field application along c-direction, perpendicular to the a-b plane of the sample. Several magneto-electronic hysteresis loops, characterized by sharp jump in R$_{xy}$, centered at non-zero field values develop at high temperature. As temperature reduces, hysteresis loops converge and eventually disappear. (d-f) Similar behavior, albeit weakly, is observed in the longitudinal resistance R$_{xx}$. (g-l) No magneto-electronic hysteresis is detected in inplane field measurements. Insets in fig. f and fig. l show measurement schematics.} \vspace{-4mm}
\end{figure}

Unlike the magnetometry study, Hall probe measurements of NiSi unveil strong magneto-electronic hysteresis curves centered around non-zero magnetic field values; thus, hinting the role of antiferromagnetic spin correlation. Hall measurements were performed for both perpendicular and inplane magnetic field applications with respect to the a-b plane of NiSi. In Fig. 4a-f, we show the plots of Hall resistance \textit{R$_{xy}$} and longitudinal resistance \textit{R$_{xx}$} as a function of perpendicular magnetic field at representative temperatures (also see fig. S5). The finite value of \textit{R}$_{xy}$ = 85 $\times 10^{-6} \: \Omega$ at \textit{H} = 0 T suggests AHE in NiSi, which can be ascribed to the uncompensated FM moment. Since the magnetic moment per lattice unit cell is not zero, the time inversion symmetry of the system is broken explicitly. It is a necessary condition for non-zero AHE.\cite{Tatara, Chang} Besides the uncompensated moment, the non-collinear antiferromagnetic ordering of Ni atoms may also be contributing to AHE via the accumulation of finite Berry curvature, as inferred previously in other AFM materials.\cite{Chang, Surgers, Smejkal} In Addition to AHE, the Hall data reveal several interesting trends: first, the system exhibits magneto-electronic hysteresis in both \textit{R$_{xy}$} and \textit{R$_{xx}$}. We observe two sets of hysteresis at \textit{T} = 250 K: one centered at low field of {$\mu_{0}$}\textit{H} $\sim$ 1.2 T, characterized by the gradual enhancement in Hall resistance as a function of the forward field sweep from {$\mu_{0}$}\textit{H} = $-$7 T to $+$7 T. When the field sweep direction is reversed, then the enhancement in \textit{R$_{xy}$} switches from positive to negative field, thus developing a magneto-electronic hysteresis. Second set of electronic hysteresis develops at higher field, marked by the sharp metamagnetic-type transitions at ${\mu_{0}}|$\textbf{\textit{H}}$| >$ 3 T. At \textit{T} = 250 K, two sets of sharp jumps are observed at higher field application. The occurrence of magneto-electronic hysteresis centered at non-zero field values suggest the AFM origin of the phenomena. An electronic hysteresis centered around $\textit{H} = 0$ T can be arising either due to FM or AFM magnetic order (as in the case of Mn$_{3}$Si), but the non-zero field center unequivocally suggests the AFM origin.\cite{Taguchi}

Second, the field-induced asymmetric jump in \textit{R$_{xy}$} (\textit{R$_{xx}$}) gradually diminishes as temperature reduces. For example, the measurement at \textit{T} = 100 K reveals the occurrence of electronic hysteresis only at higher field. At much lower temperature e.g. \textit{T} = 20 K, the hysteresis behavior completely disappears. Rather, linear magnetic field dependences are observed in both the \textit{R$_{xy}$} and \textit{R$_{xx}$} data, indicating ordinary Hall effect. These observations suggest the prominent role of thermal fluctuation in the electronic hysteresis development. Unlike the electronic behavior in perpendicular field, no magneto-electronic hysteresis is observed in \textit{R$_{xy}$} or \textit{R$_{xx}$} measurements in the inplane field, see Fig. 4g-l. In the latter case, both current and field are applied in the same direction that nullify the field effect on electric charge carriers.

\begin{figure}
\centering
\includegraphics[width=9. cm]{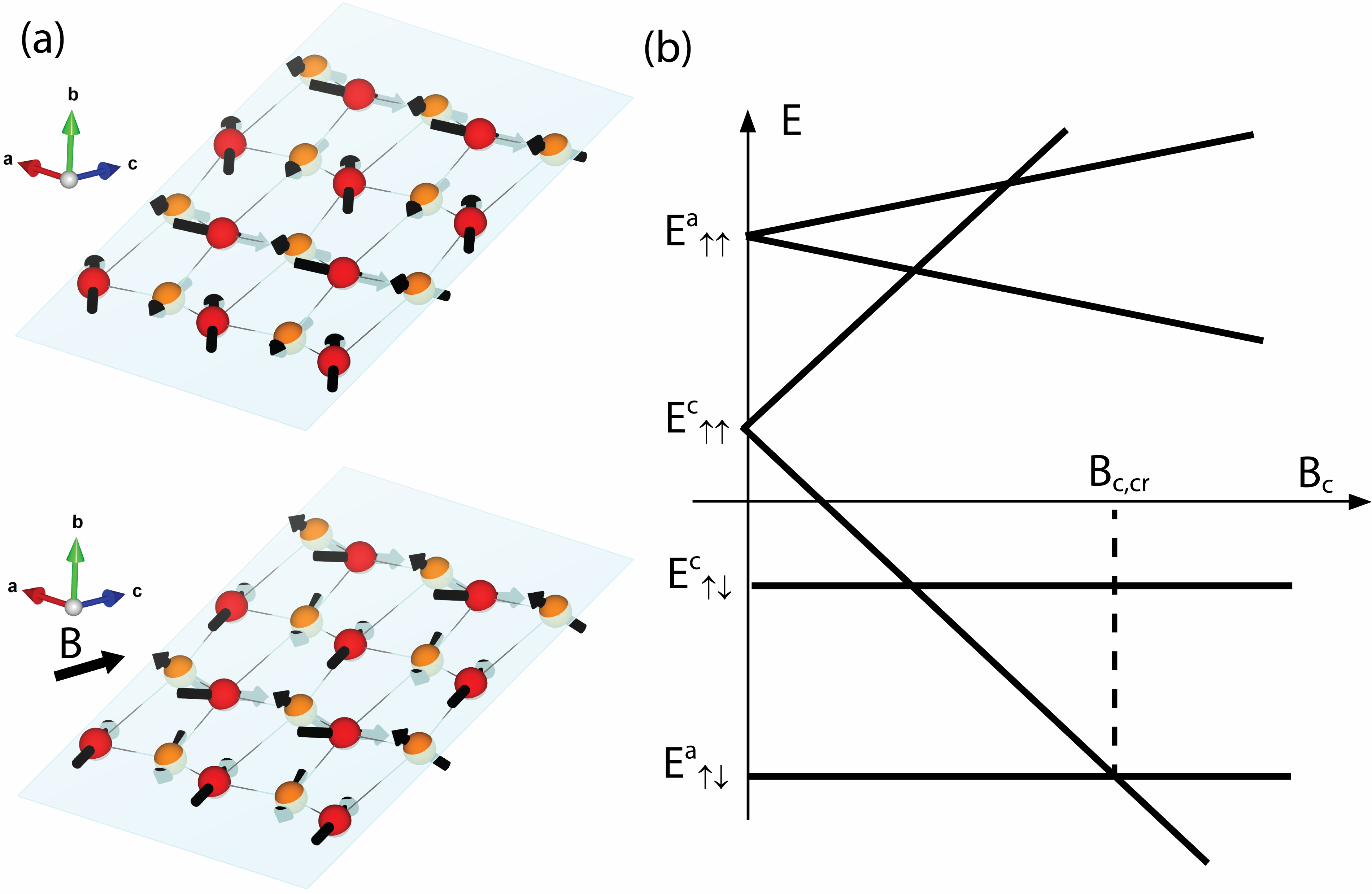} \vspace{-2mm}
\caption{\textbf{Schematic depiction of field-induced metamagnetic transition.} (a) Magnetic field application along c-axis induces torque on Ni moments, thus forces it to align along the field-direction. It results in quasi-ferromagnetic arrangement of Ni moments along c-direction, but remains antiferromagnetic along the a-direction. (b) Qualitative plot, showing energy of different magnetic configurations as a function of external field along the c-axis.} \vspace{-4mm}
\end{figure}

In principle, the magneto-electronic hysteresis in perpendicular field can be arising due to the cumulative effects of Berry curvature due to the non-collinear spin configuration, spin re-orientation and AHE.\cite{Smejkal, Taguchi} However, a straightforward energetic analysis for magnetic field application along c-direction (perpendicular to a-b plane of the sample) can provide qualitative insight into the phenomena. We can consider a model where a chain of localized moments parallel to the a-axis in (111) diagonal plane undergoes a transition from AFM to FM state, see Fig. 5a. We write the total energy of the chain as,
\begin{equation} \label{eqn}
\textit{E} = \sum_{i}\big( J_aM^a_{i}M^a_{i+1} +J_cM^c_{i}M^c_{i+1}-g{\bf M} _i\cdot {\bf B}\big)
\end{equation}
where M$_{i}$ is the total moment of chain \textit{i} and \textit{J}$_{a}$, \textit{J}$_{c}$ are AFM exchange constants along a- and c-directions, respectively (similar analysis can be performed for AFM order of total chain moment along b-direction, while magnetic field is applied along c-direction). Assuming that \textit{J}$_{a}$ $>$ \textit{J}$_{c}$ $> 0$, we find that the ground state of the system corresponds to AFM ordering of the chain moments $\textit{M}_{i}$ along the a-direction. The energy of this state is denoted by $E^{a}_{\uparrow\downarrow}$. Similarly, $E^{c}_{\uparrow\downarrow}$ corresponds to the AFM ordering of $\textit{M}_{i}$ along the c-direction. Due to the AFM ordering of chain moments, the total energy $E^{a}_{\uparrow\downarrow}$ in this configuration does not depend on magnetic field. On the other hand, energies $E^{a}_{\uparrow\uparrow}$ and $E^{c}_{\uparrow\uparrow}$, corresponding to the FM ordering of $\textit{M}_{i}$ along a- and c-directions, respectively, exhibit linear dependences on magnetic field. The qualitative plots of the energies of different magnetic configurations as a function of magnetic field are shown in Fig. 5b. We find that for magnetic field application along c-direction, a transition to FM ordering of $\textit{M}_{i}$ along c-direction with minimum energy occurs for $B_{c} > B_{cr}$. Interestingly, in-plane field application does not lead to any reconstruction of magnetic moment, which is consistent with experimental observation in Fig. 4g-l. Since the transition between different energy states of the system requires overcoming a potential barrier (see Figure S6-S7), this can lead to the hysteresis in magnetization dependent properties, such as Hall resistance. However, at low temperature, the probability of overcoming the potential barrier is smaller. Hence, the hysteretic behavior disappears.

Finally, we discuss the experimental results. The synergistic study using first principle calculations and detailed experimental investigations of single crystal specimen has revealed non-compensated antiferromagnetic ground state in NiSi. In a highly surprising observation, the antiferromagnetic ground state is onset at high temperature of $T \geq$ 700 K. There are very few AFM metals with $T_N \geq$ 500 K, in particular with non-compensated magnetization. Magnetic character of NiSi is also confirmed from bulk magnetization and the Hall probe measurements. There are both fundamental and practical implications of the finding. First of all, our study strongly complements the existing as well as the ongoing efforts on the elucidation of novel magnetic and topological ground states in bi-elemental transition metal intermetallic compounds, such as MnSi and FeSi. Second, the system manifests significant anomalous Hall conductivity at high temperature. It further validates the argument that an antiferromagnetic metal can also exhibit anomalous Hall effect.\cite{Chen} Although, the magnetic structure in NiSi has small FM component in the form of uncompensated moment, but the anomalous Hall conductivity is quite large. Therefore, it may not be entirely arising due to the latter contribution. Third, the magneto-electronic and magnetic properties at high temperature in NiSi can have implication for spintronics research. It exhibits a one-step switching transition, owed to the response of non-compensated moment to moderate magnetic field. The phenomenon is in stark resemblance with ferromagnetic materials of spintronic importance. 

Even though the uncompensated moment accompanies the AFM correlation of Ni ions, it behaves independently in perpendicular field application. This is because the AFM and uncompensated FM moments point in different directions. Thus, they respond differently to magnetic field applications. The switching property, which persists well above the room temperature and thus fulfills one of the key criteria for practical applications, can be employed for the MRAM device design. Similarly, the magneto-electronic hysteresis, attributed to sharp jumps in magnetoresistance at moderate field values, can be further explored for application in magnetic logic devices, such as memristor devices that exploits the field-induced history of magnetoresistance. \cite{Marti} Although, the phenomenon is qualitatively explained in terms of metamagnetic transition between the field-induced arrangement of Ni moments, further theoretical understanding of the origin of magneto-electronic hysteresis in NiSi is highly desirable.

\begin{figure*}
\centering
\includegraphics[width=15. cm]{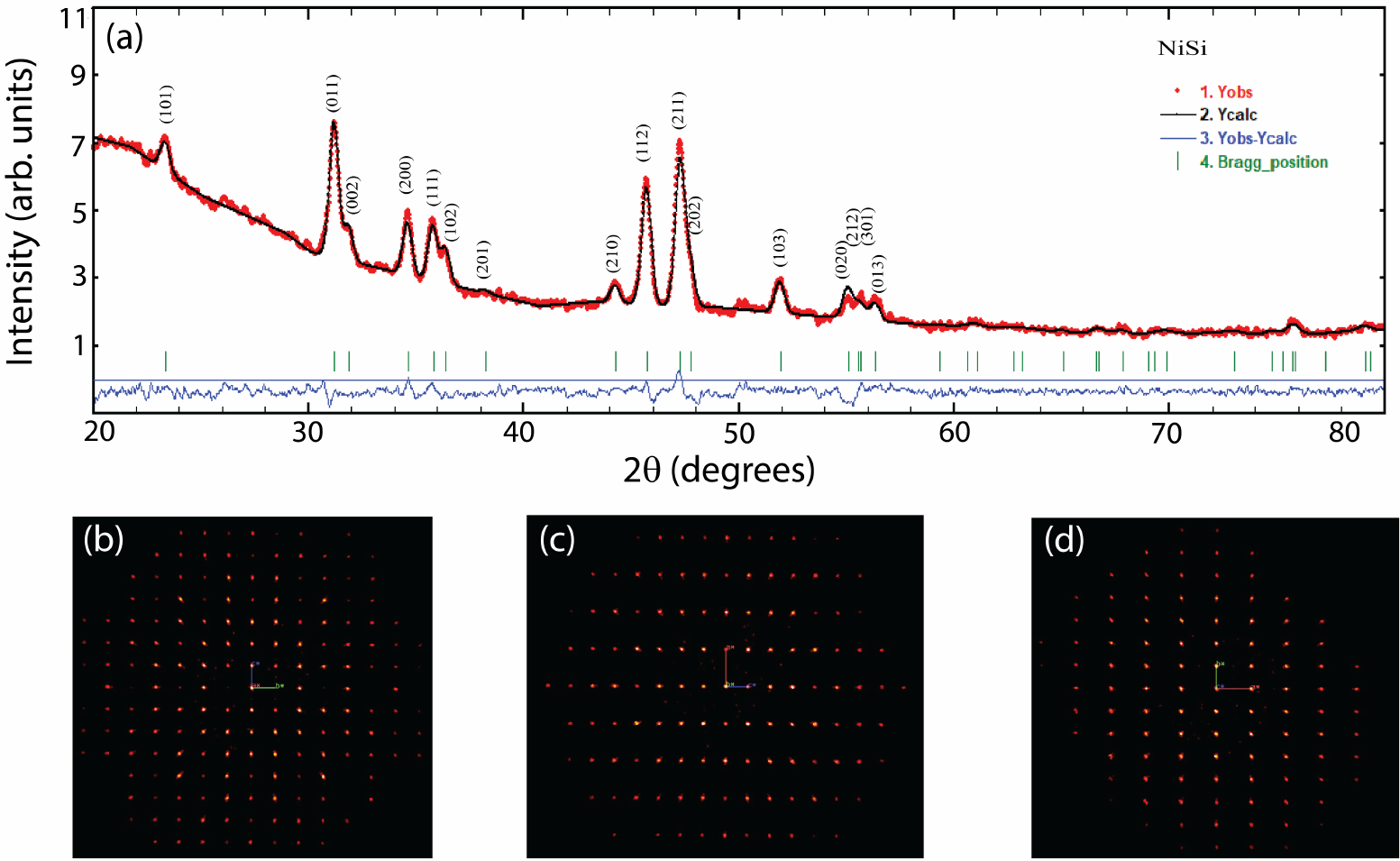} \vspace{-2mm}
\caption{(a) Powder X-ray Diffraction pattern of NiSi. Rietveld analysis of the powder diffraction data is performed using FULLPROF Suite. All detected peaks are identified with the orthorhombic lattice structure of space group pnma. (b-d) Single crystal X-ray diffraction pattern (obtained using a standard Rigaku single crystal diffractometer at SNS, ORNL) of the reciprocal lattice of NiSi at $T$ = 250K projected along the a, b, and c- direction, respectively. All detected peaks are identified with the orthorhombic lattice structure of space group pnma.} \vspace{-4mm}
\end{figure*}

\textbf{Experimental Section}

\textit{Synthesis of single crystal NiSi} - Single crystal samples of Nickel Silicide were synthesized by implementing slow cooling method with high purity powders of Ni (99.996\%) and Si (99.999\%) as starting materials from Alfa Aesar. The two powders are mixed in their stochiometric ratio (1:1) and finely ground using an agate mortar and pestle.\cite{Dahal} This mixture is then packed into a latex balloon and twisted back and forth from the outside in order to fill-in possible air gaps along the length of the balloon. Thereafter, the powder-packed balloon is evacuated and sealed before putting it in the hydrostatic press at a pressure of $\sim$ 65 MPa for 20 minutes. Finally, it is inserted in a one-side sealed quartz tube (inner diameter $=$ 7 mm) followed by evacuating the tube with the sample and sealing it on the other side. A programmable furnace is utilized to enable better control of the rate of heating and cooling. The sample is sintered at 900$^{\circ}$C for 24 hours in the programmable furnace. Subsequently, the sample is re-sintered at 1050$^{\circ}$C for 24 hours followed by cooling it at ~0.35$^{\circ}$C/min. The resulting specimen is $\sim$ 9.1mm(length) $\times \sim$ 7mm(diameter). The sample is characterized using X-ray diffraction, which confirms high chemical purity of the specimen, see Fig.6 below. FullProf refinement of XRD data confirms the zinc blend-type orthorhombic structure of NiSi with lattice parameters of a = 5.178 $\AA$, b = 3.331 $\AA$, c = 5.1624 $\AA$.

\textit{Neutron scattering measurements} - Elastic neutron scattering measurements were performed on single crystal sample of NiSi on two different spectrometers: thermal triple axis spectrometer TRIAX and elastic diffuse scattering time-of-flight spectrometer CORELLI at the University of Missouri Research Reactor and at Spallation Neutron Source at the Oak Ridge National Laboratory, respectively. Measurements on TRIAX were performed on 1.2 g single crystal sample. Experimental data were collected between $T$ ~ 5 K and $T$ = 300 K. Elastic measurements were performed at the fixed final energy of 14.7 meV using pyrolytic graphite (PG) monochromator. The measurements employed a flat PG analyzer with collimator sequence of PG filter - 60' - 60' - sample - 40' - PG filter - 40'. The sample was mounted at the end of the cold finger of a closed cycle refrigerator with a base temperature of T $\sim$ 5 K.

Higher temperature data were obtained on CORELLI spectrometer. The single crystal sample was installed inside an aluminum thermal shield and heated in a cryo-furnace in a wide temperature range from $T$ = 300 K to $T$ = 700 K. The sample was exposed to a white beam of incident energy 10 - 200 meV and a large volume of reciprocal space consisting multiple Brillouin zones were explored by rotating the sample by 360$^{o}$ in 1$^{o}$ step. Experimental data were collected at several temperatures of $T$ = 300, 550, 630 and 700 K. Data were treated with the cross-correlation method which reconstructs the elastic scattering signal.

\textit{Magnetic measurements} - Bulk magnetization measurements were performed in a SQUID magnetometer with a base temperature of $T \sim$ 5 K. Magnetic field was applied along the $c$-axis of the sample.

\textit{Hall probe measurements} - Hall measurements were performed on a $\sim$ 2.0 sq.mm NiSi specimen in van der Pauw configuration. Longitudinal resistance (R$_{xx}$) and Hall resistance (R$_{xy}$) are measured utilizing a Closed cycle refrigerator (CCR) system and Resistance Bridge. Measurements were performed for magnetic field applications perpendicular as well as in-plane (parallel to a-b plane) to the sample. Data were collected at several temperatures. At each temperature, experimental data were collected for magnetic field sweep from $H$ = 7 Tesla to -7 Tesla and from -7 Tesla to 7 Tesla. In addition to Fig. 4, additional data at $T$ = 150 K and $T$ = 50 K for perpendicular magnetic field application to the a-b plane of the sample are shown in Fig. S5.

\textbf{Acknowledgements}

PG and JG made equal contributions to this research. DKS thankfully acknowledges support from the U.S. Department of Energy, Office of Basic Energy Sciences under Grant No. DE-SC0014461. The research conducted at ORNL’s Spallation Neutron Source is supported by the Office of Basic Energy Sciences at DOE.

\textbf{Keywords}

Antiferromagnetic metal, Transition metal intermetallics, Spintronics, Neutron scattering

\clearpage

\end{document}